\documentclass[a4paper,fleqn,usenatbib]{mnras}
\usepackage{newtxtext,newtxmath}
\usepackage[T1]{fontenc}
\usepackage{ae,aecompl}
\usepackage{graphicx}	
\usepackage{amsmath}	

\newcommand{\beq}{\begin{equation}}
\newcommand{\eeq}{\end{equation}}

\def\asec{\ifmmode ^{\prime\prime}\else$^{\prime\prime}$\fi}

\def\msun{\hbox{M$_{\odot}$}}

\def\degs{\ifmmode ^{\circ}\else$^{\circ}$\fi}
\def\amin{\ifmmode ^{\prime}\else$^{\prime}$\fi}
\def\asec{\ifmmode ^{\prime\prime}\else$^{\prime\prime}$\fi}

\def\degs{\ifmmode ^{\circ}\else$^{\circ}$\fi}
\def\amin{\ifmmode ^{\prime}\else$^{\prime}$\fi}

\def\cm{\mbox{\,cm}}

\def\cm3{\mbox{\,cm$^{-3}$}}
\def\kms{\mbox{\,km~s$^{-1}$}}

\def\kms{\mbox{\,km s$^{-1}$}}

\def\lsim{\!\!\!\phantom{\le}\smash{\buildrel{}\over
 {\lower2.5dd\hbox{$\buildrel{\lower2dd\hbox{$\displaystyle<$}}\over
                                 \sim$}}}\,\,}
\def\gsim{\!\!\!\phantom{\ge}\smash{\buildrel{}\over
{\lower2.5dd\hbox{$\buildrel{\lower2dd\hbox{$\displaystyle>$}}\over
                               \sim$}}}\,\,}

\def\araa{ARA\&A}             
\def\apj{ApJ}                 
\def\apjl{ApJ}                
\def\apjs{ApJS}               
\def\apss{Ap\&SS}             
\def\aap{A\&A}                
\def\aapr{A\&A~Rev.}          
\def\mnras{MNRAS}             
\def\nat{Nature}              

\title[Free-free absorption in hot relativistic flows: application to fast radio bursts ]{Free-free absorption in hot relativistic flows: application to fast radio bursts}

\author[]{
Esha Kundu$^{1}$\thanks{E-mail: esha.kundu@gmail.com},  
Bing Zhang$^{2}$ 
\\
$^{1}$International Centre for Radio Astronomy Research, Curtin University, Bentley, WA 6102, Australia\\
$^{2}$Department of Physics and Astronomy, University of Nevada, Las Vegas, Las Vegas, NV 89154
}
\date{Accepted XXX. Received YYY; in original form ZZZ}

\pubyear{2021}

\begin{document}
\label{firstpage}
\pagerange{\pageref{firstpage}--\pageref{lastpage}}
\maketitle

\begin{abstract}
Magnetic flares create hot relativistic shocks outside the light cylinder radius of a magnetised star. Radio emission produced in such a shock or at a radius smaller than the shock undergoes free-free absorption while passing through the shocked medium. In this work, we demonstrate that this free-free absorption can lead to a negative drift in the frequency-time spectra. Whether it is related to the downward drift pattern observed in fast radio bursts (FRBs) is unclear. However, if the FRB down drifting is due to this mechanism then it will be pronounced in those shocks that have isotropic kinetic energies $\gsim10^{44}$ erg. In this model, for an internal shock with a Lorentz factor $\sim100$, the normalised drift rate $|{\rm DR_{\rm obs}}|/\nu_{\rm mean}$ is $\sim10^{-2}$~per ms, where $\nu_{\rm mean}$ is the central frequency of the radio pulses. The corresponding radius of the shocked shell is, therefore, in the range of $10^{10}$ cm and $10^{11}$ cm. This implies that, for an outflow consisting of hydrogen ion, the upper limit on the mass of the relativistic shocks is a few $\times~10^{-10}~\msun$, which is considerably low compared to that ejected from SGR\,1806-20 during the 2004 outburst.

\end{abstract}

\section{Introduction}
\label{sec:intro}

Astrophysical shocks are often considered to be associated with the most energetic events in the universe. After the death of massive stars, destruction of WDs, or mergers of compact objects, shocks are expected to be launched into the ambient medium \citep{chevalier82,Matzner99,Meszaros97}. For a normal core collapse supernova or thermonuclear runway explosions, these supernova-driven shocks are usually non-relativistic, expanding in the surrounding media with a speed $\sim$ a few $\times 10^4$ \kms \citep{chevalier16,parrent14}. Some special supernovae and compact binary coalescence of two neutron stars (NSs) or a NS and black hole can launch relativistic outflows that power energetic gamma rays bursts (GRBs) \citep{meszaros06,zhang18_grb_book}. In the case of GRBs, both internal and external relativistic shocks are possible, with the former believed to power the observed $\gamma$-ray emission \citep{Rees94} and the latter believed to power the multi-wavelength afterglow \citep{Meszaros97,Sari98}. These shocks may produce low frequency radio bursts through synchrotron maser mechanisms \citep{usov00,sagiv02,lyub14,belo17,waxman17,plotnikov19,metzger19,belo20,Yu21}. 

\par 
Fast radio bursts (FRBs) are mysterious radio transients from cosmological distances \citep{Lorimer07,Petroff19,cordes19}. It is unclear what astrophysical objects are the main sources of FRBs and whether they are generated within the magnetosphere of a highly magnetized object (e.g. a magnetar) or from relativistic shocks \citep{zhang20_nat_nov}. The very high observed event rate, e.g. more than $10^3$ $\rm sky^{-1}$ events everyday \citep{Petroff19}, suggests a high event rate density, i.e. $\sim 10^4$ $\rm Gpc^{-3} \rm yr ^{-1}$ \citep{luo20,ravi20}. One therefore is expected to detect a few of them from the nearby galaxies or even in Milky Way within a reasonable time interval. The discovery of the galactic FRB 200428 in association with the galactic magnetar SGR 1934+2154 \citep{chime20_galacticFRB,bochenek20,Mereghetti20,Li20_Xray_galSGR,ridnaia20} confirmed this and suggested that at least magnetars can make FRBs.

\par 
The immediate surrounding of the FRB sources may cause the radio wave signal to undergo several absorption and scattering processes. For example, a radio wave may suffer from free-free absorption \citep{Luan14,Murase16,yang17,kundu20}, induced Compton and Raman scattering \citep{Lyubarsky08,Lu18,pawan_kumar20,Ioka20}, and synchrotron absorption \citep{yang16} before making a way out of its production sites. Some of these conditions demand that the FRB outflow moves with a relativistic speed \citep[e.g.][]{Murase16}. In the following, we discuss what happens when a low frequency signal, produced by whatever physical mechanism, passes through a hot relativistic shell and undergoes free-free absorption in that medium. The applications and the consequences of this absorption process to FRBs are examined and discussed in $\S$ \ref{sec:app_FRB} and $\S$ \ref{sec:dis}.

\section{Free-free absorption in a hot relativistic shell}
\label{sec:ff_abs_rel_shell}
We consider two shells with a relative Lorentz factor $\Gamma_{\rm rel}$ collide and drive a pair of internal shocks. If the two shells merge after the collision and have a bulk Lorentz factor $\Gamma$, the lab-frame total energy can be estimated as
\beq
E = N  \Gamma  (\Gamma_{\rm rel} - 1) m_p c^2,
\label{eq:E_intrnl_shck_1}
\eeq
 where  $m_p$ and $c$ are the proton mass and the velocity of light in vacuum, respectively, and $N = 4 \pi n r^2 \Delta r$ is the total number of protons in the shell, $n$ is the number density of the protons in the lab-frame,  We consider two relativistic shells with a not-too-large relative Lorentz factor, so that $r = c \Gamma^2 t_{\rm obs}$ is the radius of the shock from the central engine, $\Delta r$ is the thickness of the shock in the lab-frame, and $t_{\rm obs}$ is the time in the observer frame. Combining these, we get
\beq
n = \frac{E}{4  \pi  m_p  c^4  t_{\rm obs}^2 (\Delta r)  \Gamma^5  (\Gamma_{\rm rel} - 1) },
\label{eq:E_intrnl_shck_2}
\eeq
The particle number density in the comoving frame, $n'$, is related to the density in the lab frame, $n$, through $n = n' \Gamma$. Similarly, $\Delta r = \Delta r'/\Gamma$. The frequency in the observer frame, $\nu_{\rm obs}$, is approximately equal to $2 \Gamma \nu'$. Since the particles are relativistic in the shock comoving frame, in this frame the free-free absorption coefficient can be written as 
\beq
\alpha_{\nu'}^{\rm 'ff} = 0.018  Z^2  \bar{g}'_{\rm ff}  T'^{-3/2}  n'_e  n'_i  \nu'^{-2}  (1 + A  T') 
\label{eq:ff_abs_1}
\eeq
\citep{ribicki79}, with $A = 4.4 \times 10^{-10}$ K$^{-1}$. Here $Z$ is the atomic number of the gas. $n'_e$ and $n'_i$ represent the densities of electrons and ions in the shock, respectively. Again we assume that the shell is made up of hydrogen ion. Therefore, $n'_e = n'_i (\equiv n')$ and  $Z = 1$. Also $\bar{g}'_{\rm ff}$ represents the velocity averaged Gaunt factor and $T'$ is the temperature of the electrons in the shock comoving frame. When the shocked energy is shared between electrons and protons, the temperature of the electrons behind the shock depends on the shock kinematics through the following relation
\beq
T' = \frac{\epsilon_e (m_p/m_e)}{k_B}  (\Gamma_{\rm rel} - 1)  m_e  c^2,
\label{eq:temp_elec}
\eeq
\citep{Meszaros93}, where $\epsilon_e$ is the fraction of the post shock energy that goes to electron. $k_B$ is the Boltzmann constant and $m_e$ represents electron mass. Assuming $\epsilon_e = 0.1$, one gets $T' \approx 1.1 \times 10^{12}$ K for $(\Gamma_{\rm rel} - 1) \approx 1$. Thus, $1 + A ~ T' \approx A ~ T'$. Radio pulses remain trapped in this shock until the medium is optically thick to the waves. When it becomes transparent to a given frequency the optical depth for that frequency in the shock becomes unity, i.e.,
\beq
\alpha_{\nu'}^{\rm 'ff}  \Delta r' = 1,
\label{eq:opt_depth}
\eeq
which gives $t_{\rm obs}^4 = A  T'^{-1/2} ~ \nu_{\rm obs}^{-2}  \eta_1$, with
\beq 
\eta_1 = 0.018  \frac{4}{c^4  (\Delta r)  \Gamma^9  (\Gamma_{\rm rel} - 1)^2}  \bigg(\frac{E}{4 \pi  m_p  c^2}\bigg)^2.
\label{eq:eta_1}
\eeq
Therefore,
\beq
t_{\rm obs} = \frac{A^{1/4}  T'^{-1/8}  \eta}{\sqrt{\nu_{\rm obs}}},
\label{eq:tobs2}
\eeq
where
\beq
\eta = \eta_1^{1/4}.
\label{eq:eta}
\eeq
If $t_{{\rm obs},1}$ and $t_{{\rm obs},2}$ are the times when the shell becomes transparent to $\nu_{\rm obs,1}$ and $\nu_{\rm obs,2}$, then the drift rate, $\rm DR $, can be approximated as 
\beq
{\rm DR} = \frac{\nu_{\rm obs,2} - \nu_{\rm obs,1}}{t_{{\rm obs},2} - t_{{\rm obs},1}}.
\label{eq:DR_1}
\eeq
This implies
\beq
\rm DR = -\frac{T'^{1/8}}{A^{1/4}  \eta} ~ \sqrt{\nu_{\rm obs,2} \nu_{\rm obs,1}}  (\sqrt{\nu_{\rm obs,2}} + \sqrt{\nu_{\rm obs,1}}). 
\label{eq:DR_2}
\eeq
Note that the inverse scaling between $t_{\rm obs}$ and $\nu_{\rm, obs}$ in Eq.(\ref{eq:tobs2}) stems from the $\propto \nu^{-2}$ in Eq.(\ref{eq:ff_abs_1}), which does not depend on whether the shock is relativistic, but a relativistic internal shock is needed to make the drift rate matching the observations. It is also needed to satisfy the duration and induced Compton scattering constraints if the shocks are the sites of FRBs.
For the drift rate expressed in MHz/ms, and denoted as $\rm DR_{\rm obs}$, the above equation gives
\beq
\begin{split}
\eta = -\frac{T'^{1/8}}{A^{1/4}  {\rm DR_{\rm obs}}}  \sqrt{(\nu_{\rm obs,2})_{\rm MHz} (\nu_{\rm obs,1})_{\rm MHz}} ~ \\
\Bigg[ \sqrt{(\nu_{\rm obs,2})_{\rm MHz}} + \sqrt{(\nu_{\rm obs,1})_{\rm MHz}} \Bigg],
\end{split}
\label{eq:DR_3}
\eeq
where $(\nu_{\rm obs})_{\rm MHz}$ represents frequency in MHz in the observer frame. 
Notice that when $\nu_{\rm obs,2}$ and $\nu_{\rm obs,1}$ are close to each other, their arithmetic mean would be similar to their geometric mean, so that $\eta$ is inversely proportional to the normalized drift rate ${\rm DR_{\rm obs}} / \nu_{\rm mean}$, as shown in Figure \ref{fig:eta_DR_by_numean} (discussed below).

\section{Application to Fast radio bursts}
\label{sec:app_FRB}

FRBs have been detected from 110 MHz \citep{pastor20} to 8 GHz \citep{gajjar18} with a minimum and maximum width of the pulses of around 30 $\mu$s \citep{michilli18} and 26 ms \citep{farah17} detected from FRB\,121102 and FRB\,170922, respectively. One interesting feature of FRBs, especially of those repeating ones, is the sub-pulse drifting of frequency. Low frequency subpulses are observed to be delayed with respect to high-frequency ones \citep{hessels19,amiri19,Andersen2019,fonseca20,pastor20,luo20_b}. The observed drift rates for CHIME FRBs are in the range 1 MHz/ms to 30 MHz/ms with one of the bursts from the second repeating FRB\,180814.J0422+73 having a minimum drift rate of $\sim 1$ MHz/ms. At higher frequencies, the drift rates have increased significantly as exhibited by bursts from the first repeating FRB\,121102, though some of the bursts of FRB\,180916B, detected with the Apertif telescope, displays a drift rate as small as 4 MHz/ms in the L band. 

\begin{figure}
\centering
\includegraphics[width=8.5cm,origin=c]{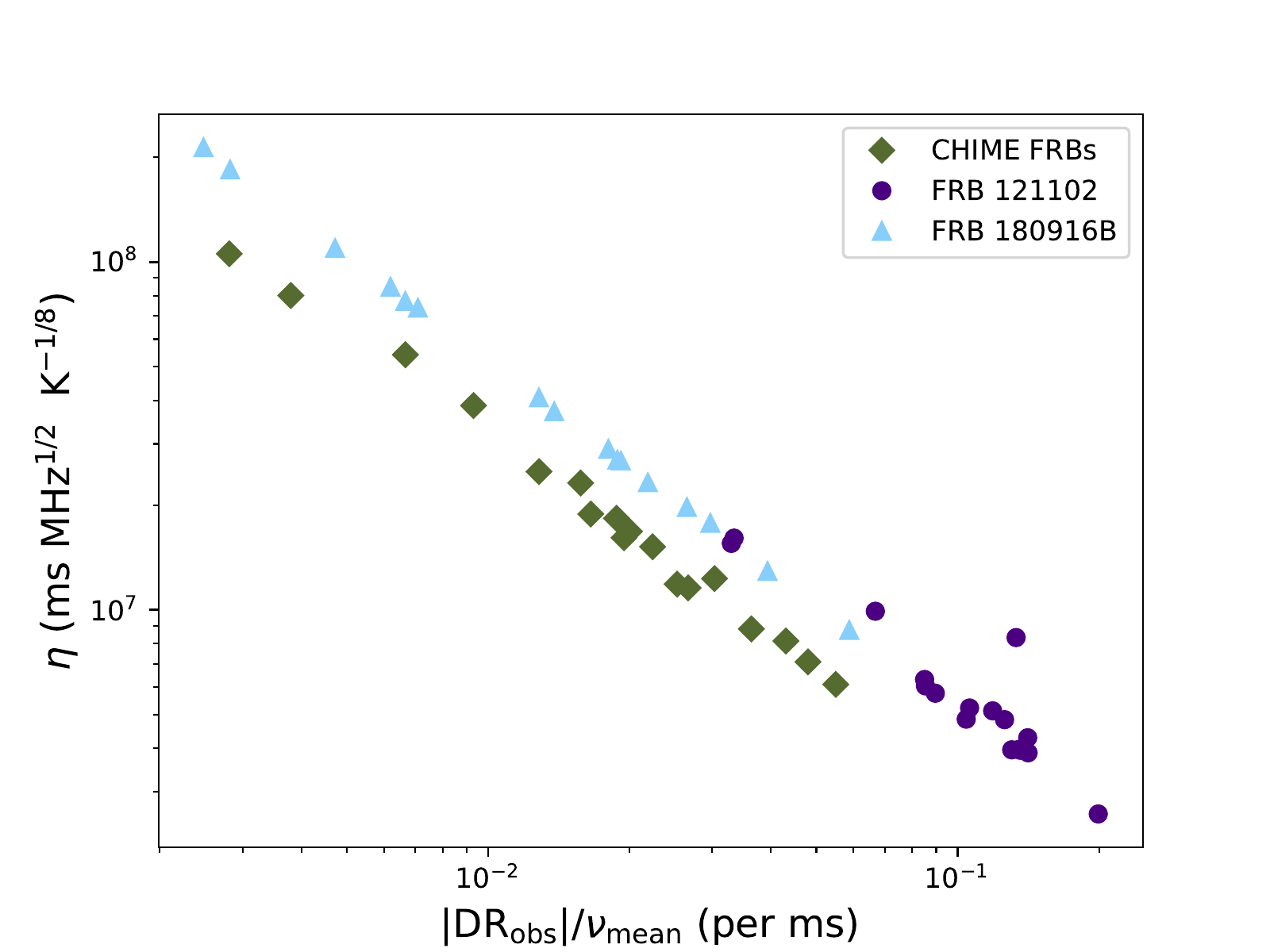}
\caption{The required $\eta$ as a function of modulus of $\rm DR_{\rm obs}$ divided by the central frequency of the observed pulse for an internal shock. The $\rm DR_{\rm obs}$ of different FRBs are taken from     \citet{hessels19,Andersen2019,fonseca20,pastor20}. The filled diamonds, triangles and circles represent the CHIME repeaters, FRB\,121102 and FRB\,180916B, respectively. The CHIME sample includes the second repeating FRB\,180814.J0422+73.}
\label{fig:eta_DR_by_numean}
\end{figure}

We apply the free-free absorption theory discussed in Section \ref{sec:ff_abs_rel_shell} to FRBs and to investigate whether it can interpret the down-drifting feature. Since FRBs are millisecond duration transient radio pulses, one can take $\Delta t_{\rm obs} \sim 1$ ms. The millisecond duration of the bursts implies that the characteristic length scale of the emission region is $\Delta r \sim c \Delta t_{\rm obs} \sim 3\times 10^7$ cm. Inserting this value of $\Delta r$ in Eq. \ref{eq:eta_1} and using Eq.\ref{eq:eta} we obtain
\beq
\Gamma^9  (\Gamma_{\rm rel} - 1)^2 = \frac{8.3 \times 10^{42}}{\eta^4} E_{\rm 45}^2,
\label{eq:eq21}
\eeq
where $E_{\rm 45} =E (\rm erg)/10^{45}$ erg. Over the frequency range of 400 MHz to 7 GHz, the observed drift rates, $\rm DR_{\rm obs}$, of different bursts vary from $\sim - 1$ MHz/ms to $- 870$ MHz/ms. The emission bandwidth of most of the bursts is small: for the CHIME bursts it is around $200$ MHz. From Eq.\ref{eq:eta}, the estimated values of $\eta$ for an internal shock are in the range of $2 \times 10^{6}$ to $2 \times 10^{8}$ ms MHz$^{1/2}$ K$^{-1/8}$, where $T' = 1.1 \times 10^{12}$ K. Figure \ref{fig:eta_DR_by_numean} displays the required $\eta$ as a function of $|{\rm DR_{\rm obs}}|/\nu_{\rm mean}$ with $\nu_{\rm mean} = (\nu_{\rm obs,2} + \nu_{\rm obs,1})/2$, which represents the central frequency of the observed radio pulse. The values of the DR of different FRBs were published in    \citet{hessels19,Andersen2019,fonseca20,pastor20}. The filled diamonds, triangles and circles represent the CHIME repeaters, FRB\,121102 and FRB\,180916B, respectively, where the second repeating FRB\,180814.J0422+73 is included in the CHIME sample. 

\par
\begin{figure}
\centering
\includegraphics[width=8.5cm,origin=c]{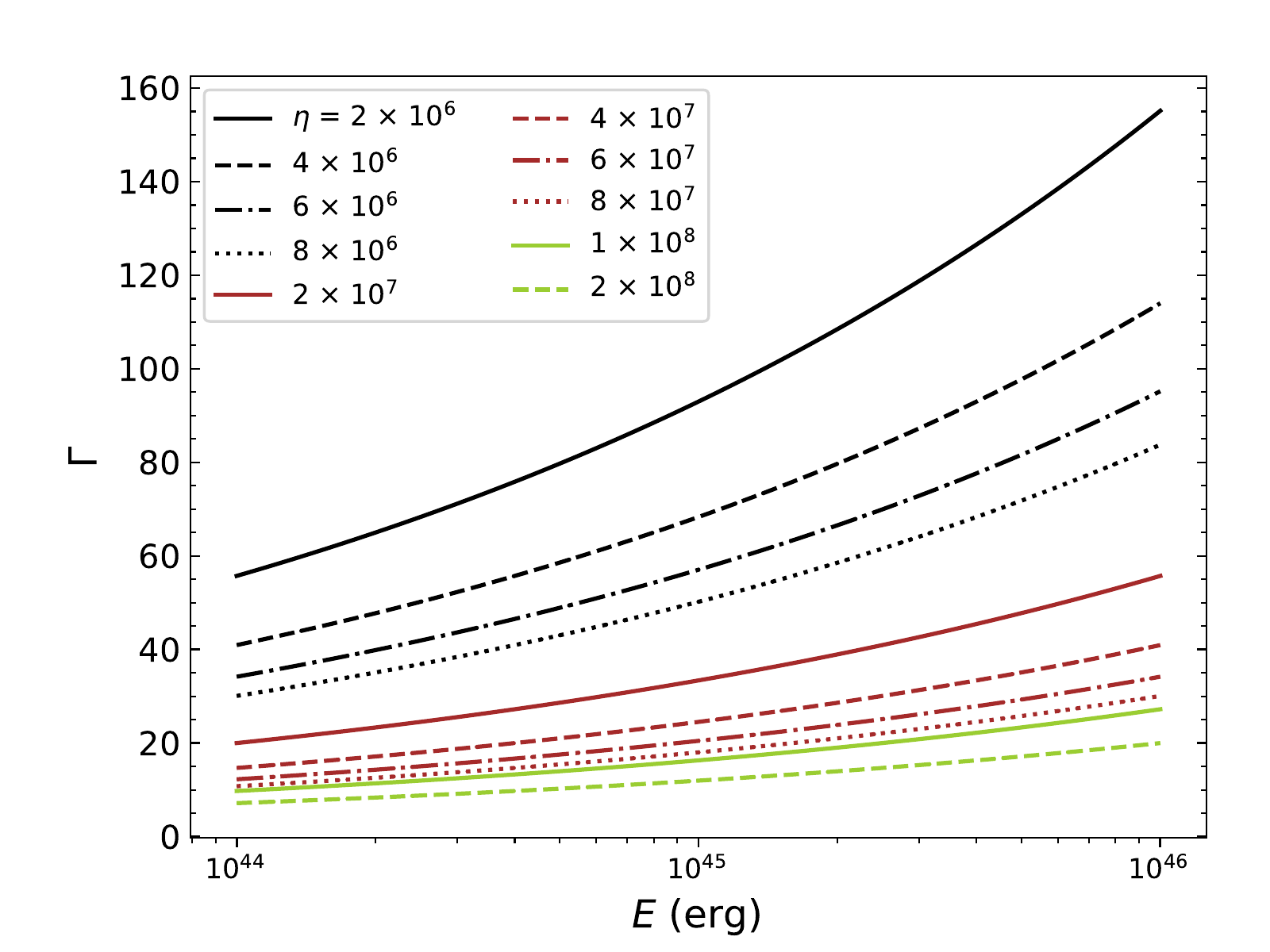}
\caption{The allowed values of $\Gamma$ as a function of shock energy, $E$, for $\eta$ in the range $2 \times 10^{6}$ to $2 \times 10^{8}$ ms MHz$^{1/2}$ K$^{-1/8}$.}
\label{fig:gamma_E}
\end{figure}

\begin{figure}
\centering
\includegraphics[width=8.5cm,origin=c]{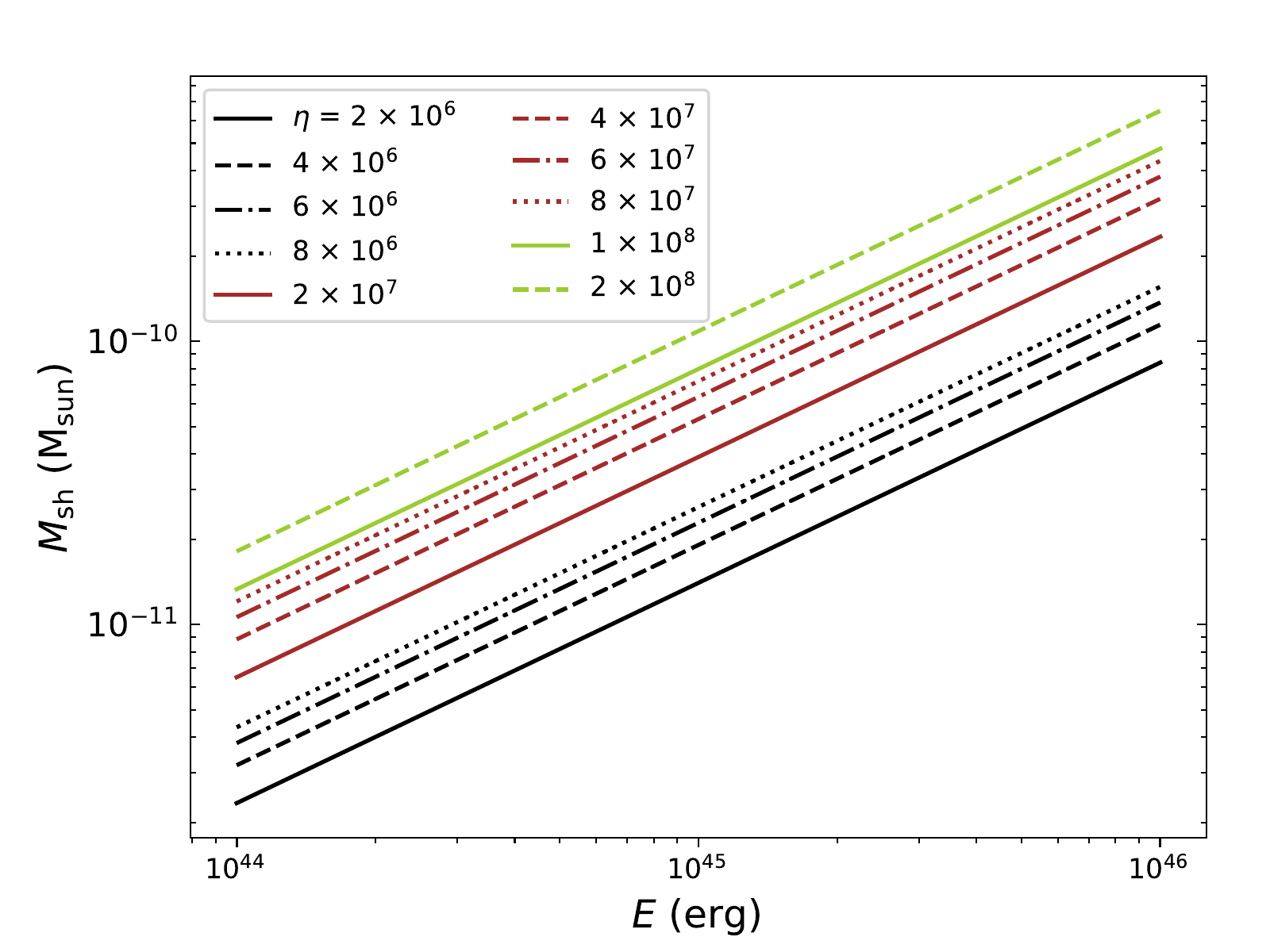}
\caption{The mass of the shell, $M_{\rm sh}$, as a function of shock energy $E$ for different values of $\eta$ obtained for an internal shock.}
\label{fig:E_mass}
\end{figure}

 
The magnetic energy stored in a magnetar with a radius $\sim$ 10 km and a characteristic magnetic field $B \sim 10^{15}$ G \citep{duncan_RC92} is $\gsim 10^{47}$ erg. For $B \gsim 10^{16}$ G the dipolar magnetic energy could be as high as $10^{49}$ erg. The giant flare detected from SGR\,1806-20 in 2004, during a hyperactive phase, had an isotropic flare energy of around $2 \times 10^{46}$ erg \citep{palmer05}. Moreover, it was found that a similar amount of energy was released from a magneter in NGC 253 during an extremely bright gamma-ray burst event on 15th April 2020 \citep{Svinkin21}.  
For $\eta$ in the range from $2 \times 10^{6}$ to $2 \times 10^{8}$ ms MHz$^{1/2}$ K$^{-1/8}$, the allowed values of $\Gamma$ (estimated using eq.\ref{eq:eq21}) as a function of the shock energy $E$ varying from $10^{44}$ erg to $10^{46}$ erg are shown in Fig.\ref{fig:gamma_E}. We note that free-free absorption is pronounced for shocks having higher kinetic energies, $\gtrsim 10^{44}$ erg (otherwise, $\Gamma$ is too small to satisfy the drift-rate constraint). 
The particle density in the shocks is
\beq 
n = \frac{2.4 \times 10^{-25}  \eta^4  \Gamma^4  (\Gamma_{\rm rel} - 1)}{t_{\rm obs}^2  E_{\rm 45}},
\label{eq:eq22}
\eeq
as obtained from Eqs.\ref{eq:E_intrnl_shck_2} and \ref{eq:eq21}. Assuming that shocks are solely made up of hydrogen and $t_{\rm obs} \sim 1$ ms, the mass of the shells, $M_{\rm sh}$, as a function of $E$ is depicted in Fig.\ref{fig:E_mass} for different values of $\eta$, estimated for an internal shock. For $E \sim 10^{44}$ erg, $M_{\rm sh}$ is around $10^{-12}$ $\msun$; and it is about two orders of magnitude higher when the shock energy is $10^{46}$ erg\footnote{ For the giant flare detected from SGR\,1806-20, \citet{Granot06} invoked a mildly relativistic outflow to interpret the radio afterglow and derived a lower limit on the ejecta mass as $\ge 1.5 \times 10^{-9}$ \msun. If FRBs are related to shocks with that set of parameters, then the down-drifting rate cannot be interpreted within the model proposed here.}. For an active magnetar, the maximum mass of the ejected material during a flaring event could be $\sim 10^{-5}$ \msun \citep{belo17}. In our model, even when the highest energy outbursts occur everyday, a magnetar would require around 100 yr to eject a total of $\sim 10^{-5}$ \msun  matter in the surrounding medium.

\par
\begin{figure}
\centering
\includegraphics[width=8.5cm,origin=c]{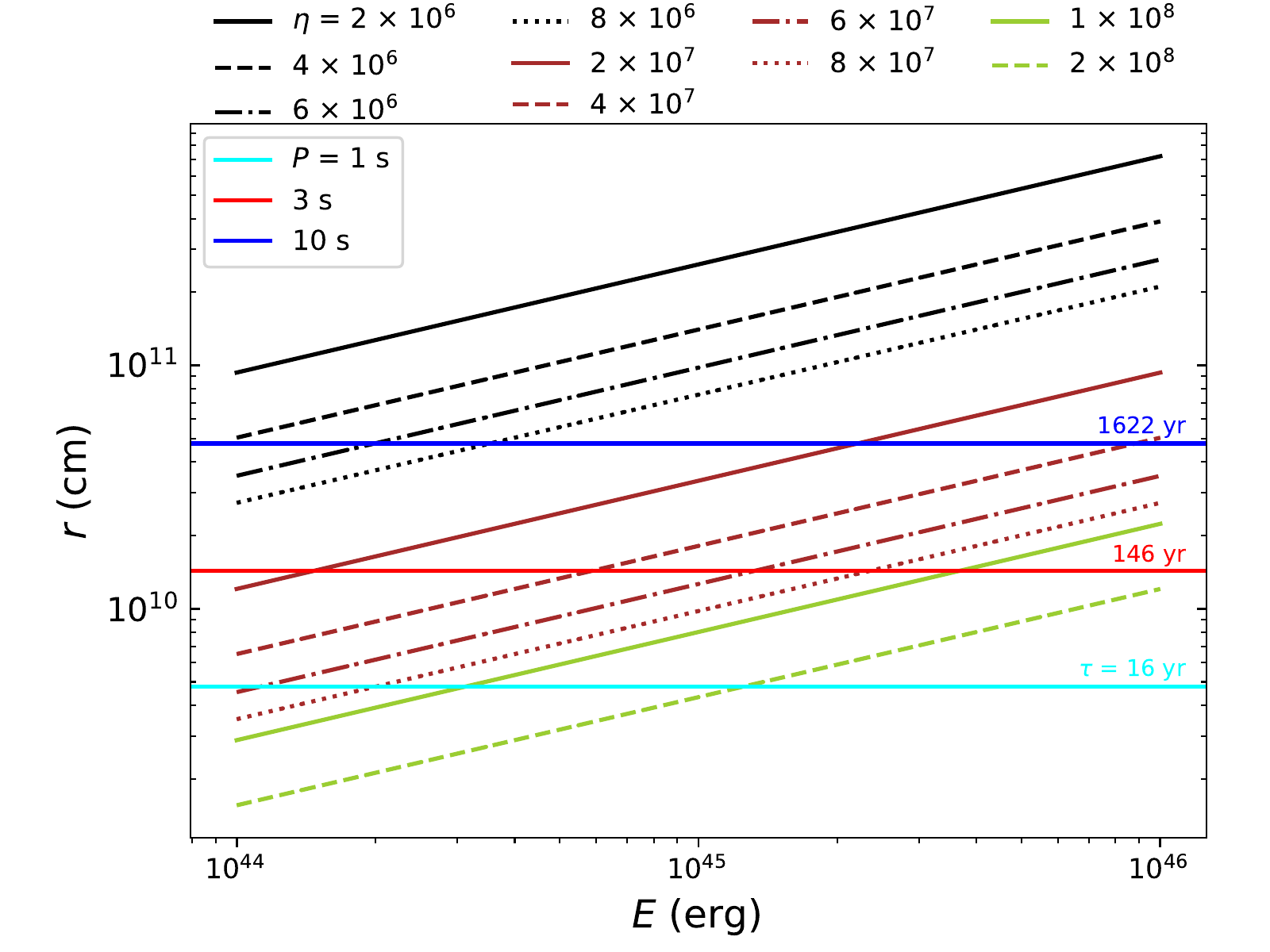}
\caption{Radius of the shell as a function of  shock energy $E$ for different values of $\eta$. The cyan, red and blue horizontal lines display the radius of the LC when the period $P$ of a magnetar is 1 s, 3 s and 10 s, respectively. The age of the magnetar corresponding to each $P$ is shown in the plot, assuming that the spindown energy loss is dominated by magnetic dipole radiation.}
\label{fig:E_radius}
\end{figure}

The radius of the shell as a function of the shock energy $E$ for different values of $\eta$ is displayed in Fig.\ref{fig:E_radius}. Since relativistic shocks are expected to form beyond the magnetosphere, we also draw the magnetar light cylinder (LC) radius $R_{\rm LC} \simeq c  P/(2\pi)$ (which defines the outer boundary of the magnetosphere) for comparison, where $P$ is the magnetar spin period. 
In the figure, the cyan, red and blue horizontal lines exhibit the radii of the LC for $P = $  1 s, 3 s and 10 s, respectively. Assuming that the spindown energy loss of a magnetar is dominated by magnetic dipole radiation, the characteristic age of the magnetar can be estimated as $\tau \simeq {P}/(2 \dot{P})$, where the braking index of the magnetar is presumed to be 2, and $\dot{P}$ represents the derivative of the period. For a magnetar of mass $1.4 ~ \msun$ and radius 10 km, an upper limit on $\dot{P}$ is calculated as $\dot{P} ({\rm {s \ s^{-1}}}) = [B ({\rm {G}})/(3 \times10^{19})]^2/{P (\rm {s})}$. Considering $B = 10^{15}$ G, the age $\tau$ of the magnetar is found to be around 16 yr, 146 yr and 1622 yr for $P = $  1 s, 3 s and 10 s, respectively. This implies that the older the population, the smaller the $\eta$ values for free-free absorption to be a relevant process to produce the observed down drifting in the frequency-time plane.

\section{Discussion} 
\label{sec:dis}

FRBs from magnetars have been interpreted within the framework of two types of models. One group of models suggests that coherent emission originates from the magnetosphere of the neutron star \citep{kumar17,yang18,yang21,Lu18,Lu20}, while the other group of models interprets coherent emission as synchrotron maser in relativistic shocks \citep{lyub14,belo17,waxman17,plotnikov19,metzger19,belo20}. Within the magnetosphere model, the frequency down-drifting has been interpreted as a consequence of ``radius-to-frequency mapping'' commonly observed in radio pulsars, i.e. emission with different frequencies originate from different altitudes and the high-frequency emission is observed earlier than low-frequency one \citep{wang19,Lyutikov20}. Within the synchrotron maser model, the drift is interpreted as due to emitting frequency decreasing with radius in relativistic shocks \citep{metzger19,belo20}. Polarization angle swings of some FRBs favor the magnetosphere origin of FRBs \citep{luo20_b}, but it is unclear whether the shock model may be responsible for some FRBs.

\par 
In this work, we illustrate that the free-free absorption in relativistic shocks introduces a feature of down-drifting. Whether it is related to the FRB phenomenology is unclear. 
However, if the FRB down drifting is due to this mechanism then the following conditions should be satisfied: i.) there should be a hot relativistic shell through which the radio waves pass through and ii.) the energy of this shock should be $\gsim 10^{44}$ erg. In this model, the shell serves as the absorber of the radio waves. The radiation may originate from the shock, similar to what happens in the case of synchrotron maser model of FRBs, or could be produced at a radius smaller than the hot shell, e.g. from the magnetosphere of the central engine\footnote{For the latter scenario, the radio waves encounter the shock while propagating out so that a downdrifting pattern can be observed only when the shock just turns from optically thick to optically thin in the observing frequency band during the encounter. Since not all the bursts from repeating FRBs have the down-drifting pattern, according to our interpretation, those with such a pattern are the ones that satisfy this requirement.}. 
These shocks are expected to have $\Gamma$ much greater than unity. In the case of an internal shock, this criteria is fulfilled except for $\eta \sim  10^{8}$ ms MHz$^{1/2}$ K$^{-1/8}$ and $E < 10^{45}$ erg.  The majority of the bursts (see Fig.\ref{fig:eta_DR_by_numean}) require $\eta < 4 \times 10^{7}$ ms MHz$^{1/2}$ K$^{-1/8}$. For $\eta \sim  10^{6}$ ms MHz$^{1/2}$ K$^{-1/8}$ and $E \sim 10^{45}$ erg, $\Gamma$ is $\sim 100$ (see Fig.\ref{fig:gamma_E}). The smaller values of the $\eta$ imply larger shock radii (see Fig.\ref{fig:E_radius}). For the shocks with a radius $\sim 10^9$ cm from the central engine,  the magnetar that powers FRBs could be as young as a couple of decade old, as demonstrated in Fig.\ref{fig:E_radius} based on the assumption that the spin down energy loss is dominated by the magnetic dipole radiation. Furthermore, the mass of the shocked shell is less than a few $\times 10^{-10}$ $\msun$ as shown in Fig.\ref{fig:E_mass}. This suggests that our model may work even for a magnetar that is active for $\gsim 100$ yr and flares almost every day. 

\par 
 Particle-in-cell simulations of synchrotron maser emission from relativistic shocks suggests that the efficiency of this mechanism is $7 \times 10^{-4}/\sigma^2$, for $\sigma \gsim 1$, where $\sigma$ is the magnetisation parameter \citep{plotnikov19}. This implies that the efficiency of synchrotron maser is very low requiring a large amount of energy to go to other wavelengths. The observed radio-to-X-ray flux ratio for FRB 200428 associated with the 
 Galactic magneter SGR 1934+2154 \citep{Mereghetti20,Li20_Xray_galSGR,ridnaia20,tavani_nature20,chime20_galacticFRB,bochenek20} roughly satifies this constraint \citep{Margalit20_galFRB}, even though the magnetoshere model can also interpret this ratio \citep{Lu20,yang21}. The energy of the localised cosmological FRBs are in the range $10^{27} - 10^{34}$ erg/Hz \citep{tendulkar17,bannister19,Ravi19_nature,marcote20_nature,law20_nature,bhandari20,bhandari20_FRB191001}. Using the dispersion measure-redshift \citep{deng14} relation \citet{zhang_2018_nonlocFRB_lum_eng} demonstrates that the peak luminosities of the FRBs are in the range $10^{42} - 10^{45}$ erg $\rm {s}^{-1}$. These imply that the radio energy of the cosmological FRBs vary between $10^{36} - 10^{43}$ erg. Even though no simultaneous X-ray bursts has been detected from cosmological FRBs despite attempts \citep{Scholz16}, if one assumes that the X-ray-to-radio luminosity ratio of these FRBs is similar to that of FRB 200428, one would expect that the total energy of these events is at least $10^{40} - 10^{47}$ erg. For $E>10^{44}$ erg free-free absorption in  shocks would be important in producing the downward drift in frequency-time as demonstrated in Fig.\ref{fig:gamma_E}. As a result, the mechanism discussed in this paper should be considered in FRB modeling and may be relevant to the down-drifting feature observed in at least some FRBs. 

\section*{Acknowledgements}
E.K acknowledges the Australian Research Council (ARC) grant DP180100857.  

\section*{Data Availability}
The data underlying this article will be shared on reasonable request to the corresponding author.

\bibliographystyle{mnras}
\bibliography{referns}

\bsp	
\label{lastpage}
\end{document}